\documentclass[aps,prb,twocolumn,
	groupedaddress,superscriptaddress,
	amsfonts,amssymb,amsmath,floatfix,
	citeautoscript]{revtex4-1}

\usepackage{graphicx}
\usepackage[centering,hmargin=20mm,tmargin=30mm,bmargin=25mm]{geometry}
\usepackage{multirow}
\usepackage{newtxtext}
\usepackage[cmintegrals]{newtxmath}

\usepackage{xcolor}
\usepackage{hyperref}
\hypersetup{colorlinks,
	linkcolor={blue!75!black!80!yellow},
	citecolor={blue!75!black!80!yellow},
	urlcolor={blue!75!black!80!yellow}
}

\makeatletter
\renewcommand\@make@capt@title[2]{%
	\@ifx@empty\float@link{\@firstofone}{\expandafter\href\expandafter{\float@link}}%
	\sffamily{\textbf{#1}}\@caption@fignum@sep#2
}%

\makeatother

\renewcommand{\vec}[1]{\textbf{#1}}
\renewcommand{\Im}{\operatorname{Im}}
\renewcommand{\Re}{\operatorname{Re}}
\newcommand{\Tr}{\operatorname{Tr}}
\newcommand{\D}{\mathrm{d}}
\newcommand{\sub}[1]{\ensuremath{_{\textrm{#1}}}}   
\newcommand{\iu}{\mathrm{i}}
\newcommand{\appropto}{\mathrel{\vcenter{
	\offinterlineskip\halign{\hfil$##$\cr
	\propto\cr\noalign{\kern.2pt}\sim\cr\noalign{\kern-2.5pt}}}}} 

\usepackage{xspace} 
\newcommand{\ie}{i.e.\@\xspace} 
\newcommand{\cf}{cf.\@\xspace}
\newcommand{\eg}{e.g.\@\xspace}

\newcommand{\RPIMSE}{Department of Materials Science and Engineering, Rensselaer Polytechnic Institute, Troy, NY, USA}
\newcommand{\UCdavis}{Department of Chemistry, University of California, Davis, CA, USA}
\newcommand{\HarvardSEAS}{John A. Paulson School of Engineering and Applied Sciences, Harvard University, Cambridge, MA, USA}
\newcommand{\MITPhy}{Department of Physics, Massachusetts Institute of Technology, Cambridge, MA, USA}

\renewcommand{\textmu}[1]{\ensuremath{\mu}}

\begin{document}

\title{Plasmonics in Argentene}

\author{Ravishankar Sundararaman}\email{sundar@rpi.edu}\affiliation{\RPIMSE}
\author{Thomas Christensen}\affiliation{\MITPhy}
\author{Yuan Ping}\affiliation{\UCdavis}
\author{Nicholas Rivera}\affiliation{\MITPhy}\affiliation{\HarvardSEAS}
\author{John D. Joannopoulos}\affiliation{\MITPhy}
\author{Marin Solja\v{c}i\'{c}}\affiliation{\MITPhy}
\author{Prineha Narang}\email{prineha@seas.harvard.edu}\affiliation{\HarvardSEAS}

\date{\today}

\begin{abstract}
Two-dimensional materials exhibit a fascinating range of electronic and
photonic properties vital for nanophotonics, quantum optics and emerging 
quantum information technologies. Merging concepts from the fields of
\emph{ab initio} materials science and nanophotonics, there is now an opportunity to engineer new photonic materials whose optical, 
transport, and scattering properties are tailored to 
attain thermodynamic and quantum limits.
Here, we present first-principles calculations predicting that \emph{Argentene},
a single-crystalline hexagonal close-packed monolayer of Ag, can dramatically
surpass the optical properties and electrical conductivity of conventional plasmonic materials.
In the low-frequency limit, we show that the scattering rate and resistivity
reduce by a factor of three compared to the bulk three-dimensional metal.
Most importantly, the low scattering rate extends to optical frequencies
in sharp contrast to e.g.\ graphene, whose scattering rate increase drastically in the
near-infrared range due to optical-phonon scattering.
Combined with an intrinsically high carrier density, this facilitates
highly-confined surface plasmons extending to visible frequencies.
We evaluate Argentene across three distinct figures of merit, spanning the spectrum of 
typical plasmonic applications; in each, Argentene outperforms the state-of-the-art. 
This unique combination of properties will make Argentene a valuable addition to the 
two-dimensional heterostructure toolkit for quantum electronic and photonic technologies.
\end{abstract}

\maketitle

Two rapidly developing and converging fields from the past decade make atomic-scale engineering of new materials now within reach. First, a revolution in materials discovery has yielded a diverse portfolio
of new classical and quantum photonic materials. These include a variety of nanostructures and two-dimensional layered architectures that can be crafted with structural precision approaching the
atomic scale. Secondly, advances in nanophotonics, plasmonics, and metasurfaces have enabled
precise control of light-matter interactions down to the nanoscale. Frontiers in the science of new materials increasingly focus on novel phenomena and properties that emerge in the limit of extreme
quantum confinement and low dimensionality. Reduction of three-dimensional (3D) materials to their two-dimensional (2D) equivalents, a single atomic layer, results in qualitatively different properties compared to bulk or even few-layered materials. Well-known examples include graphite exhibiting Dirac points in its monolayer equivalent, graphene, and the indirect to direct band-gap transition in bulk vis-\`a-vis monolayer MoS\sub{2}. Further, the range of possible optical and electronic phenomena is especially rich
in these materials due to interlayer coupling in 2D material heterostructures.
For example, optical properties can be tailored and exotic states of matter created by altering the 
layering sequence~\cite{Novoselov:2005, Gjerding:2016,Andersen:2015} or twist angle between layers~\cite{Herrero:2018a,Herrero:2018b}. 

2D conductors like graphene are particularly interesting for their unique optical properties~\cite{Jablan:2013, Bludov:2013, Abajo:2014, Stauber:2014, Low:2014, Low:2016, Basov:2016, AdvOptMat, Papadakis:2018fk}. The surface plasmon resonance of 2D and ultra-thin conductors exhibits
a drastically different plasmon dispersion relation from bulk 3D conductors,
with an order of magnitude higher mode confinement.\cite{PRB_2018_PNEK}
Consequently, 2D materials are expected to introduce a paradigm shift
in photonics and optoelectronics, condensing optical phenomena
to the few nanometer scale and enabling strong interaction between  
quantum emitter and plasmons.\cite{Alcaraz-Iranzo:2018zl, Mertens:2017uq}
However, the low intrinsic carrier densities and strong optical-phonon
scattering in known 2D conductors so far limit the regime of 
low-loss 2D plasmonics to mid-infrared frequencies, approximately
one order of magnitude below the visible spectrum.\cite{Ni:2018fv}

Delivering the promise of 2D plasmonics and nanophotonics\cite{Pendry:1999db, KhurginConfinement} to the visible region while retaining low loss and long propagation lengths, requires true 2D metals with carrier densities two orders of magnitude higher than present-day 2D conductors
(which are doped semi-metals) and semiconductors, and without optical-phonon losses.
Model calculations of single-layer Ag and Au, that treat the hypothetical 2D metal
as a 2D electron gas at the jellium level~\cite{Manjavacas:2014},
or simply as a conductive sheet with properties extrapolated from its bulk
dielectric function~\cite{Jablan:2013, Abajo:2015}, 
already show tremendous potential for this class of materials.
However, there are two fundamental limitations with such model calculations.
First, the all-important scattering time that determines loss in the material
is unknown and treated as an empirical parameter, at best extrapolated from
its the bulk value, while scattering times in deposited thin films of metals
decrease with film thickness.\cite{Fuchs,Sondheimer,SurfaceScattering}
Second, and more importantly, it is unclear from previous model calculations
if the material would remain stable in its monolayer form.

Our work overcomes these fundamental limitations in literature to calculate a new class of monolayer
plasmonic metals. Here, we use \emph{ab initio} calculations based on density-functional theory
to show that a monolayer of Ag atoms can form a stable
2D hexagonal close-packed lattice, which we henceforth refer to as Argentene.
Furthermore, from first-principles electron-phonon scattering calculations,
we predict that the momentum relaxation time in single-crystalline Argentene not only matches
the value of perfect bulk Ag, but that it in fact exceeds it by a factor of three!
Correspondingly, the conductivity of Argentene is three times larger than
bulk Ag and is comparable to the best-case optimally-doped values for graphene.
Finally, we show that Argentene particularly shines in its optical response, 
because the absence of optical-phonon scattering allows the high relaxation time
to persist to high frequencies, unlike in graphene where it sharply drops off
past 0.2\,--\,0.5~eV photon energies.
With these \emph{ab initio} dielectric functions, we show that Argentene
exhibits low-loss highly-confined plasmons extending to the visible regime.

The experimental verification of the properties predicted here requires the identification of suitable synthesis and deposition techniques and possibly stabilizing substrates for reliable, single-crystalline growth of noble-metal monolayers. At least two points are encouraging in this regard: 
(1)~we predict that free-standing Argentene is thermodynamically stable with a $\text{0.16~eV}$ barrier, and 
(2)~previous work~\cite{Schaffner:1998vl, Ma:1995jt, Chen:2010sf, Huang:2013gd, McPeak:2015, Nagpal:2009cs} on epitaxial growth and stabilization of noble-metal nanostructures already support the practical feasibility of fabrication.

\begin{figure}[ht!]
\includegraphics[width=.925\columnwidth]{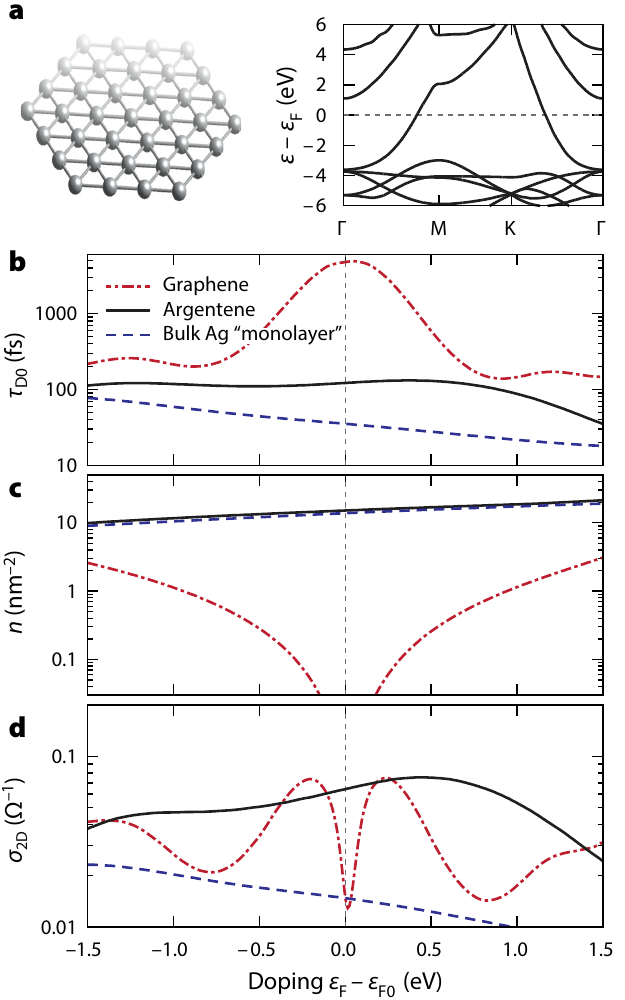}
\caption{{\bfseries\boldmath Structure and DC carrier transport in Argentene:}
\textbf{a}, Argentene is a single hexagonal-close packed atomic layer of Ag with
a 2D electron-gas-like band structure extending till the \emph{d}-bands 3.5~eV below the Fermi level.
\textbf{b}, Electron-phonon momentum relaxation time for DC transport, $\tau_{\mathrm{D}0}$ in Argentene
is three times larger than bulk Ag, and nominally independent of the Fermi level position,
and is comparable to that of heavy ideally-doped graphene (no dopant scattering).
\textbf{c}, Argentene's carrier density is an order of magnitude greater than graphene at practical doping levels,
resulting in \textbf{d}, a larger 2D conductivity through most of the relevant range.
(For comparison, results for bulk Ag in \textbf{c} and \textbf{d} are normalized
to a thickness equal to its (111)-layer separation $t\sub{2D} \approx \text{2.36~\AA}$.)
Furthermore, Argentene and bulk Ag do not require doping to conduct; the effect
of Fermi level modification in these materials is only shown for consistency.
\label{fig:bandstructTauDensity}}
\end{figure}

Argentene, a single close-packed atomic layer of Ag atoms, exhibits the band structure
of a nearly-perfect 2D electron gas for electrons near the Fermi level, as shown in
our density-functional theory (DFT) calculations in Fig.~\ref{fig:bandstructTauDensity}a.
Its quadratic dispersion relation is disrupted by $d$-bands that
start 3.5~eV below the Fermi level, remarkably similar to 3D bulk Ag.

Charge transport in Argentene is, however, markedly different from bulk Ag.
We find the electron-phonon scattering time, which critically determines
electrical conductivity and plasmonic quality factors, to be a factor of three
larger in Argentene as shown in Fig.~\ref{fig:bandstructTauDensity}b.
Note that conventional expectations from charge transport in thin \emph{imperfect}
films of noble metals that scattering time decreases with film thickness
is due to surface and grain boundary scattering.\cite{Fuchs,Sondheimer,SurfaceScattering}
These non-ideal effects are highly sensitive to growth techniques, substrates,
and over-layers; here we focus on the potential of the ideal material
and consistently compare results for perfect single crystals in both the 2D and 3D cases.

In fact, a more interesting comparison for Argentene is the best-known 2D conductor, graphene.
Note that throughout the rest of the Article we compare and contrast Argentene and graphene for consistency, though we recognize that graphene plasmonics has well-known limitations.
Perfect undoped graphene exhibits scattering time exceeding picoseconds (Fig.~\ref{fig:bandstructTauDensity}b),
but a very low carrier density (Fig.~\ref{fig:bandstructTauDensity}c)
and hence only a modest 2D conductivity (Fig.~\ref{fig:bandstructTauDensity}d).
Making graphene into a useful conductor or plasmonic material requires doping
to increase the carrier density, but this also increases the density of states
at the Fermi level and the phase space for electron-phonon scattering,
resulting in a reduction in scattering time with increasing carrier concentration.
This results in a peak 2D conductivity of 0.06~$\Omega^{-1}$ at an optimal doping level
that corresponds to a Fermi level 0.3~eV away from the Dirac point and a carrier
density $\sim 0.1$~nm$^{-2}$ $= 10^{13}$~cm$^{-2}$.
Note that for a fair comparison with single-crystal Argentene,
we consider ideal doping in graphene neglecting dopant scattering
in order to represent the best-case scenario for graphene.
Argentene matches this best-case 2D conductivity of
0.06~$\Omega^{-1}$ without need for doping.

For comparison, we also show predictions for Argentene and bulk Ag if its Fermi level
were changed by doping and find that its properties are virtually unchanged.
The scattering time is roughly constant with change of Fermi level,
consistent with the flat density of states, and hence phase space
for electron phonon scattering, of a 2D free-electron system.
(The reduction near 1~eV increase of Fermi level arises from the unoccupied band that starts
1~eV above the Fermi level in the band structure shown in Fig.~\ref{fig:bandstructTauDensity}a.)
The scattering time decreases with increasing Fermi level in bulk Ag
due to $g(\varepsilon) \propto \sqrt{\varepsilon}$ for a 3D free-electron system,
while in graphene, it decreases as the Fermi level moves away from the
Dirac point (at energy $\varepsilon_0$) due to increase in the 
density of states as $g(\varepsilon) \propto |\varepsilon-\varepsilon_0|$.
We re-iterate that Argentene does not require doping since it is a true 2D metal,
whereas graphene is a semi-metal, and all subsequent results focus on undoped Argentene.
Similarly, for comparison, we focus on undoped bulk Ag and graphene at
its best-case ideal doping of 0.3\,--\,0.5~eV in the remainder of this work.

\begin{figure}
\includegraphics[width=.925\columnwidth]{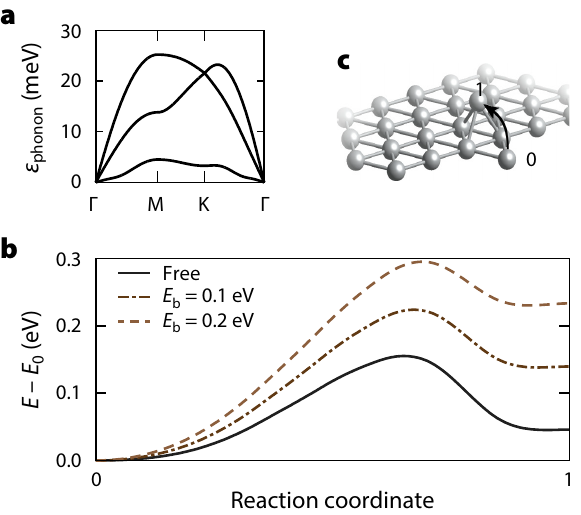}
\caption{{\bfseries\boldmath Stability of Argentene from first-principles:}
\textbf{a}, Phonon bandstructure without imaginary frequencies indicate a mechanically stable 2D layer, even when free-standing.
\textbf{b}, Kinetic stability towards island formation, evaluated using the barrier for an atom
in-plane to jump on top of the 2D layer; path is sketched in \textbf{c}.
The 0.16~eV barrier for free-standing Argentene increases as the binding energy per atom $E_{\text{b}}$
to a hypothetical van der Waals substrate increases, allowing the single atomic layer to be
further stabilized on an appropriately chosen substrate.
\label{fig:stability}}
\end{figure}

Next, \emph{ab initio} DFT calculations show that Argentene is mechanically stable as a free-standing 2D material, as indicated by the absence of any imaginary frequencies in the phonon band structure
in Fig.~\ref{fig:stability}a.
Fig.~\ref{fig:stability}b reinforces this stability by showing the barrier for an atom
in the plane of Argentene to hop onto the next layer, as sketched in Fig.~\ref{fig:stability}c,
which would be the process by which a single monolayer could transform to
islands of multiple layers or nanoparticles.
We find a kinetic barrier of 0.16~eV ($\approx 6k_{\mathrm{B}}T$ at room temperature)
for free-standing Argentene; this can be doubled when bound to a van der Waals
substrate (eg. hexagonal boron nitride) with a modest binding energy per atom $\sim 0.2$~eV.
Calculations of Argentene's stability on specific substrates and possible growth mechanisms are
subjects of ongoing work, and we focus here next on the remarkable plasmonic properties of Argentene.

Transitioning from DC and low-frequency transport properties to optical and plasmonic response of metals, the relevant material response function is the frequency-dependent complex conductivity
(closely related to the dielectric function via $\sigma(\omega) = -\iu\omega(\epsilon(\omega)-\epsilon_0)$),
which can be written as
\begin{equation}
\sigma(\omega) = \frac{\sigma_0\tau_{\mathrm{D}0}^{-1}}{\tau_{\mathrm{D}}^{-1}(\omega) + \iu\omega} + \sigma_{\mathrm{d}}(\omega),
\end{equation}
where $\sigma_0$ and $\tau_{\mathrm{D}0}$ are the DC conductivity and Drude momentum-relaxation time,
$\tau_\mathrm{D}(\omega)$ is the frequency-dependent momentum relaxation time that
encapsulates intraband phonon-assisted contributions to the optical response
and $\sigma_{\mathrm{d}}(\omega)$ is the contribution due to direct optical transitions.
We emphasize all of these quantities are calculated from a fully first-principles treatment
of electrons and phonons, explicitly including all bands, modes, and coupling
matrix elements, as discussed in the Methods section.
For 2D materials, we consider the corresponding 2D conductivities ($\sigma\sub{2D}$) rather than the bulk conductivities ($\sigma$).

\begin{figure}
\includegraphics[width=.925\columnwidth]{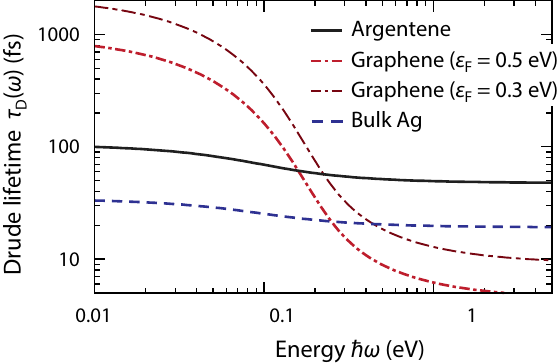}
\caption{{\bfseries\boldmath Frequency-dependence of momentum-relaxation time:}
The electron-phonon momentum (Drude) relaxation time $\tau_{\text{D}}(\omega)$ of graphene (Fermi levels $\varepsilon_{\text{F}} =\text{0.3 and 0.5~eV}$) is initially substantially
than Ag and Argentene in the low-frequency limit,
but drops dramatically at frequencies above 0.2~eV, falling below that of Argentene and Ag,
due to strong optical-phonon scattering in graphene.
Argentene's relaxation time is consistently three times larger than bulk Ag; 
both exhibit only minor reduction with increasing frequency due to the absence of an analogous optical-phonon scattering mechanism in these materials.
\label{fig:tauAC}}
\end{figure}

The frequency-dependent relaxation time $\tau_{\mathrm{D}}(\omega)$ directly determines
the intraband loss, which along with interband losses in $\sigma_{\mathrm{d}}(\omega)$,
limit the plasmonic performance.
Figure~\ref{fig:tauAC} shows that graphene's unparalleled DC relaxation time
drops by two orders of magnitude over the $\hbar\omega\sim \text{0.2\,--\,0.5~eV}$ 
frequency range, arising from scattering with optical phonons
with a maximum energy $\sim \text{0.2~eV}$.
In contrast, both bulk Ag and Argentene do not have optical phonons
and show a much more modest reduction in their relaxation times,
around a factor of two, from DC to optical frequencies.
This leads to a cross-over at $\hbar\omega \sim 0.2$~V,
where Argentene takes over as the lower-loss material.
This low-loss regime persists well into the visible region till
the interband threshold $\sim 3.5$~eV in both Argentene and Ag,
beyond which direct transitions generate high losses.

\begin{figure}
	\includegraphics[width=0.925\columnwidth]{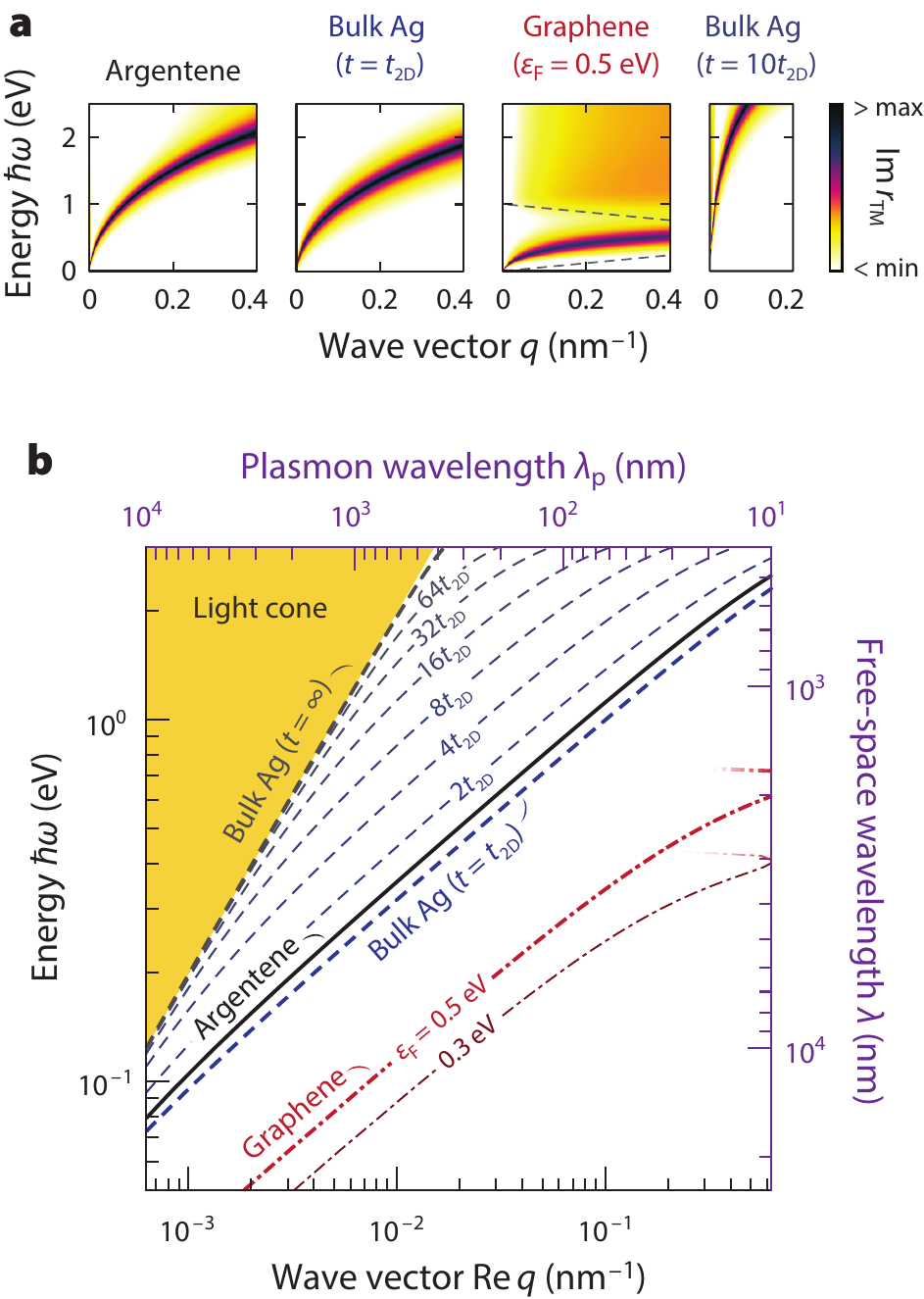}
	\caption{{\bfseries\boldmath Plasmon dispersion of Argentene, doped graphene 
			($\varepsilon_{\text{F}} = \text{0.3 and 0.5~eV}$), and thin slabs of bulk Ag:}
		\textbf{a}, Imaginary part of the TM reflectivity $\Im r\sub{tm}$
		(logarithmic, clamped colorscale) whose peaks reflect the existence of plasmon modes.
		\textbf{b}, Corresponding plasmon dispersion (solved for complex $q$ and real $\omega$)
		for Argentene, graphene, and bulk Ag slabs (thicknesses range over
		$t=2^nt_{\text{2D}}$ for $n=0,1,\ldots,6$ and $\infty$).
		The confinement of Argentene plasmons compare well with that predicted
		from extrapolation of bulk Ag properties down to a monolayer's thickness.
		In thicker slabs, however, confinement is order of magnitudes lower.
		Graphene hosts highly confined plasmon in the mid-infrared spectrum;
		Argentene's plasmons extend into the near-infrared and above.
		\label{fig:dispersion}}
\end{figure}

The plasmon dispersion of a given 2D layer is directly related to the layer's frequency-dependent 2D conductivity $\sigma\sub{2D}(\omega)$. 
Specifically, the in-plane plasmon wave vector $q$ disperses with frequency as $q = [(2\iu\varepsilon_0\omega/\sigma\sub{2D})^2 + k_0^2]^{1/2}$ (free-space wave vector, $k_0\equiv \omega/c$), reducing to $q \simeq 2\iu\varepsilon_0\omega/\sigma\sub{2D}$ in the quasistatic limit~\cite{Koppens:2011, Bludov:2013, ChristensenThesis:2017}. 
Figure~\ref{fig:dispersion} depicts the plasmon dispersion of of Argentene, doped graphene, and nanometric slabs of bulk Ag of thickness $t$ (spanning integer-multiples of Ag's (111)-layer separation $t\sub{2D}\approx \text{2.36~\AA}$). 
The plasmon's dispersion coincides with the peaks of the imaginary part of the transverse-magnetic (TM) reflection coefficient (Fig.~\ref{fig:dispersion}a); the associated peak width relates directly with the plasmon lifetime and propagation length.

At small excitation energies, the 2D layers exhibit the well-known $\omega \appropto q^{1/2}$ dispersion. 
This furnishes them with substantially larger wave vectors (at fixed frequency)---and hence stronger confinement---than their finite-thickness slab counterparts (Fig.~\ref{fig:dispersion}b). 
Given the manifold opportunities facilitated by high confinement,
the attraction of the monolayer limit is manifest: confinement is more than an order of magnitude larger in Argentene than the 16 layer Ag slab. 
The enhancement is immediately appreciable from a small-thickness analysis of the slab's dispersion equation~\cite{Economou:1969},
which demonstrates that, classically, $q(\omega)\appropto 1/t$ for $t\ll k_0 \ll |q|$.
Coincidentally, the  dispersion $\Re q(\omega)$ of the Ag slab of thickness $t=t_{\text{2D}}$,
\ie the ``monolayer'' bulk Ag slab (henceforth, ML-Ag), exhibits a perhaps counter-intuitively good agreement with Argentene. 
This, however, is to be expected: for a 2D carrier density $n$, the plasmon dispersion is $\Re q \appropto \omega^2/n^{s}$ 
in the Drude regime (with $s = 1$ in metals and $s=1/2$ in graphene, \cf its Dirac dispersion)~\cite{ChristensenThesis:2017}.
Accordingly, the observed agreement merely reflects the approximate equality of $n$ in Argentene and $n\sub{3D}t\sub{2D}$ in ML-Ag.
Argentene distinguishes itself from graphene in two ways: (1) its plasmon frequencies exceed graphene's significantly,
extending into the near-infrared and above, and (2) its confinement is smaller at equal frequencies.
Since $\Re q \appropto 1/n^{s}$, both differences are consequences of Argentene's higher carrier density.

\begin{figure}
\includegraphics[width=0.95\columnwidth]{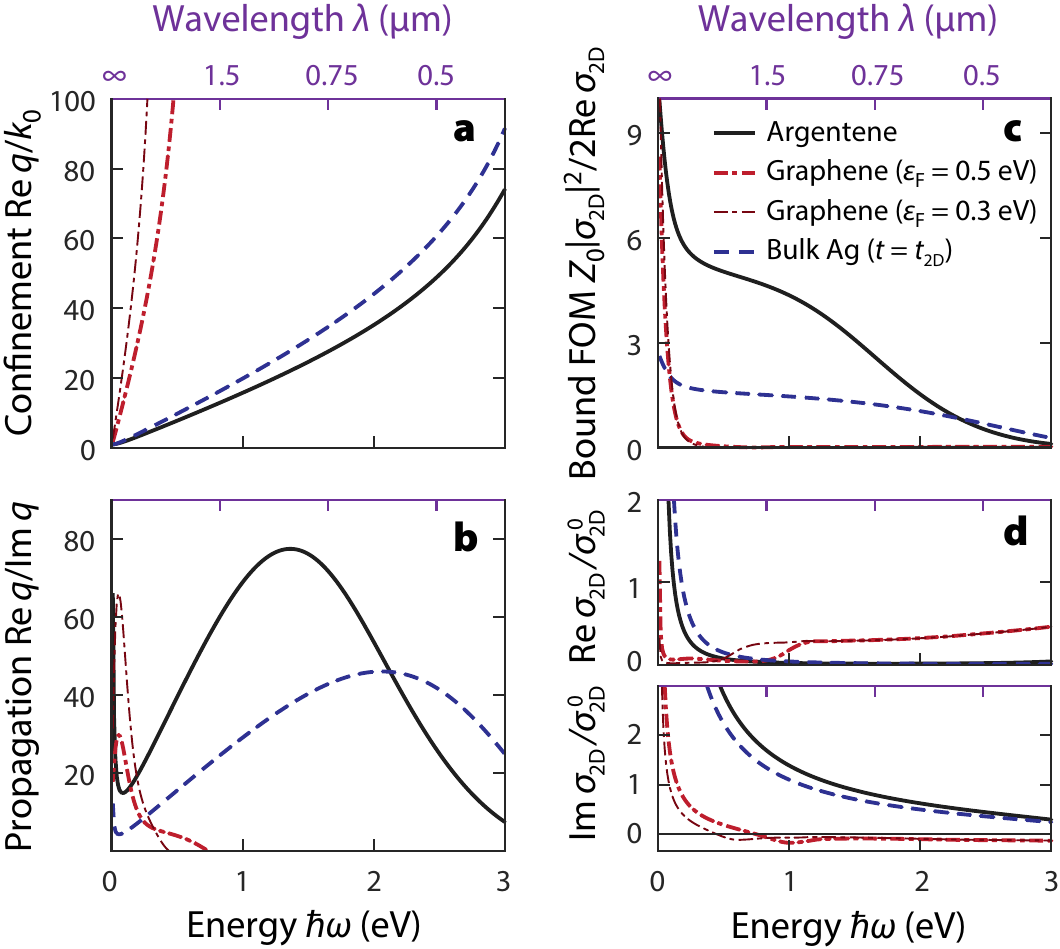}
\caption{{\bfseries\boldmath Plasmonic figures of merit in Argentene, doped graphene ($\varepsilon_{\text{F}} = \text{0.3 and 0.5~eV}$), and ``monolayer'' bulk Ag:}  
\textbf{a}, Confinement ratio $\Re q/k_0$.
\textbf{b}, Effective propagation length $\Re q/\Im q$.
\textbf{c}, Bound-related FOM $\Omega\equiv  Z_0|\sigma\sub{2D}|^2/2\Re\sigma\sub{2D}$.
\textbf{d}, Real and imaginary parts of the 2D conductivity $\sigma\sub{2D}$ (in units of $\sigma\sub{2D}^0 \equiv e^2/\hbar$).
Argentene offers roughly an order of magnitude increase in maximal effective propagation length over graphene --
whose response above 0.2~eV (below 6~\textmu{}m) is dominated by electron-phonon interaction with its optical phonon branch --
similar confinement ratios, and superior bound-FOM. Argentene's plasmonic properties are optimal near the 1.55~\textmu{}m telecommunication band.
Relative to bulk-extrapolated monolayer properties, \ie to bulk Ag slabs of thickness $t=t\sub{2D}$,
Argentene exhibits anomalously improved plasmonic attributes.
\label{fig:fom}}
\end{figure}

The ultimate merits of a given plasmonic material depend on use-case. 
In Fig.~\ref{fig:fom}a-c, we consider three distinct figures of merit (FOMs),
spanning the gamut of typical plasmonic applications: confinement ratio $\Re q/k_0$,
effective propagation length $\Re q/\Im q$, and a bound-related FOM
$\Omega \equiv Z_0|\sigma\sub{2D}|^2/2\Re\sigma\sub{2D}$ ($Z_0$, impedance of free space)~\cite{Miller:2017}. 
The latter FOM warrants further explication than the former two, which are well-established plasmonic FOMs:
$\Omega$ bounds the optical response of arbitrarily shaped 2D resonators---\eg, the extinction efficiency is $\leq 2\Omega$,
the Purcell factor is $\leq \tfrac{3}{4}(k_0d)^{-4}\Omega$, and the radiative heat flux (between identical bodies)
relative to the black-body limit is $\leq 6(k_0 d)^{-4}\Omega^2$, for emitter--body and body--body separations $d$.
In the quasistatic limit, the bound-related FOM is $\Omega \simeq k_0/\Im q$, \ie a complementary effective propagation length, taken relative to its free-space wavelength. 
Interestingly, the two conventional FOMs, confinement ratio and effective propagation length,
are similarly simply related to the conductivity in the quasistatic limit:
$\Re q/\Im q\simeq  \Im \sigma\sub{2D}/\Re \sigma\sub{2D}$ and $\Re q/k_0\simeq 2\Im\sigma\sub{2D}/Z_0|\sigma\sub{2D}|^2$. 
Thus, each FOM convey, approximately, distinct ratios of the complex components of the conductivity (Fig.~\ref{fig:fom}d).

Despite these commonalities, the three FOMs individually present contrasting pictures.
In terms of confinement (Fig.~\ref{fig:fom}a) doped graphene surpasses Argentene, while Argentene and ML-Ag agree well. 
Conversely, graphene's propagation ratios (Fig.~\ref{fig:fom}b) fall short of Argentene's, except in the low-frequency region ($\lesssim \text{0.2~eV}$). 
Similarly, the propagation ratios of Argentene and ML-Ag exhibit significant discrepancies. 
Analogous observations are evident for the bound-related FOM (Fig.~\ref{fig:fom}c).
This FOM-dependent contrast between Argentene and ML-Ag conclusions reflects a fundamental difference in the essential dependence of each FOM: 
confinement, on one hand, is a comparatively simple theoretical construct, depending mainly on macroscopic properties, specifically the carrier density $n$, as discussed previously.
On the other hand, propagation ratios (and the bound-related FOM) sensitively depend on relaxation mechanisms, which are intrinsically of a microscopic nature. 
Specifically, relaxation can occur either through direct transitions%
~\footnote{Our calculations neglect beyond-local response mechanisms, such as surface-enhanced Landau damping.
Their impact on the charge-symmetric mode is negligible~\cite{Christensen:2017},
except at higher wave vectors beyond those considered here.}
or through electron-phonon interaction. 
The latter is incorporated here via a frequency-dependent relaxation time $\tau(\omega)$,
computed from the Eliashberg spectral function.
The discrepancy between ML-Ag and Argentene underscores the need for full, microscopic accounts of 
the electron-phonon interaction in the quantitative assessment of novel 2D plasmonic materials.

The preceding discussion also explains the differences noted between graphene and Argentene: 
graphene's confinement exceeds Argentene's due to its lower carrier density, at the cost of lower operation frequencies.
In contrast, the operation range of graphene's plasmons is further restricted in practice, however, due to the onset of strong electron-phonon interaction with graphene's optical phonon branch at 0.2~eV~\cite{Jablan:2009,Jablan:2013}. 
At room-temperature, this interaction significantly broadens graphene's plasmons, near and above the threshold
(at cryogenic temperatures, strong relaxation is thresholded to energies $\gtrsim${0.2~eV}, with $\Im q$ decreased markedly below).
Argentene, a single-atom Bravais lattice, doesn't support optical phonons and consequently isn't similarly impacted.
Jointly, the three FOMs of Fig.~\ref{fig:fom}a-c underscore the appeal of Argentene for plasmonics,
the importance of microscopic accounts in theoretical assessments of novel plasmonic materials,
and the advances attainable by pursuing a still deeper pool of plasmonic platforms.


In summary, our first-principles calculations reveal that Argentene, 
a single hexagonal close-packed atomic-layer of Ag,
is mechanically stable in free-standing form.
Perfect Argentene crystals will exhibit three times the
momentum relaxation time and conductivity as bulk Ag,
roughly comparable to the best-case scenario for ideally-doped graphene.
While graphene's long scattering time and low loss regime
are limited to frequencies with $\hbar\omega \lesssim \text{0.2~eV}$
due to optical-phonon scattering, Argentene's low loss
regime extends well into the visible spectrum till an
interband threshold $\sim \text{3.5~eV}$.
Consequently, Argentene exhibits highly-confined plasmons
with long propagation lengths at much higher frequencies.
Realizing the promise of ultra-confined, long-lived, visible-spectrum 2D plasmonics
with Argentene, requires the identification of suitable substrates
and techniques to reliably grow single crystals of noble-metal monolayers,
while simultaneously retaining its superior electron-phonon scattering properties.

\section*{Acknowledgments}
The authors thank Professors Efthimios Kaxiras (Harvard University), John Pendry (Imperial College, London), Ling Lu (Chinese Academy of Science), Toh-Ming Liu (RPI), and Daniel Gall (RPI) for fruitful discussions on 2D plasmonic materials, potential monolayer growth techniques, and carrier scattering properties.
RS acknowledges start-up funding from the Department of Materials Science and Engineering at Rensselaer Polytechnic Institute. 
TC acknowledges support from the Danish Council for Independent Research (Grant No.\ DFF--6108-00667).
NR was supported by Department of Energy Fellowship DE-FG02-97ER2530 (DOE CSGF).
The research of JDJ and MS was supported as part of the Army Research Office through the Institute for Soldier Nanotechnologies under contract no.~W911NF-18-2-0048 (photon management for developing nuclear-TPV and fuel-TPV mm-scale-systems), and also supported as part of the S3TEC, an Energy Frontier Research Center funded by the US Department of Energy under grant no.~DE-SC0001299 (for fundamental photon transport related to solar TPVs and solar-TEs).
PN acknowledges start-up funding from the Harvard John A. Paulson School of Engineering and Applied Sciences. 

This research used resources of the National Energy Research Scientific Computing Center,
a DOE Office of Science User Facility supported by the Office of Science of the U.S. Department of Energy under Contract No. DE-AC02-05CH11231, 
the Research Computing Group at Harvard University as well as resources at the Center for Computing Innovations (CCI) at Rensselaer Polytechnic Institute.
\vspace{-1.5em}

\section*{Author Information}
The authors declare no competing financial interests.

\section*{Methods}

\subsection*{Computational details}
We perform first-principles calculations of electrons, phonons and their matrix elements
in the open-source JDFTx software,\cite{JDFTx} using norm-conserving pseudopotentials\cite{SG15}
at a kinetic energy cutoff of 30 Hartrees, the Perdew-Burke-Ernzerhof generalized gradient approximation\cite{PBE}
to the exchange-correlation functional and truncated Coulomb interactions
to isolate periodic images for the 2D systems.\cite{TruncatedEXX}
We use the rotationally-invariant DFT+$U$ formulation\cite{DFTplusU}
with $U = \text{2.45~eV}$ for Ag to obtain the correct $d$-band positions.\cite{DirectTransitions}
We use 24 $\vec{k}$-points along each periodic direction for Brillouin zone integration,
along with Fermi-Dirac smearing with width 0.01~Hartrees for Fermi surface sampling
in the DFT calculations.
Phonon calculations employ a $4\times 4\times 4$ supercell for bulk Ag
and $6 \times 6 \times 1$ supercell for Argentene and graphene.
All electronic and phononic properties are converted to a maximally-localized
Wannier function basis,\cite{MLWFmetal} and then interpolated to extremely fine $\vec{k}$
and $\vec{q}$ meshes ($\sim 1000$ points along each periodic direction) for all subsequent perturbation
theory calculations outlined below for optical response and carrier scattering properties.
These subsequent calculations employ electron and phonon occupation factors
at room temperature, 298~K (with $k_{\mathrm{B}}T \sim 0.00094$ Hartrees).

\subsection*{Conductivity and DC transport}

We evaluate the low-frequency conductivity using a full-band relaxation time
approximation to the linearized Boltzmann equation,\cite{PhononAssisted, Coulter:2018, NitrideCarriers}
\begin{equation}
\boldsymbol{\sigma} = \int\sub{BZ}\frac{e^2g\sub{s}\,\D\vec{k}}{(2\pi)^d} \sum_n
	\frac{\partial f_{\vec{k}n}}{\partial \varepsilon_{\vec{k}n}}
    (\vec{v}_{\vec{k}n}\otimes\vec{v}_{\vec{k}n})
    \tau^p_{\vec{k}n},
\label{eqn:sigma}
\end{equation}
where $\varepsilon$, $f$ and $\vec{v}$ are the energies, Fermi occupations,
and velocities of electrons with wave-vector $\vec{k}$ in band $n$,
and $g\sub{s} = 2$ is the spin-degeneracy factor.
The above expression automatically evaluates to the 3D conductivity $\boldsymbol{\sigma}$
for bulk Ag with $d=3$, while it is the 2D conductivity $\sigma\sub{2D}$
for Argentene and graphene with $d=2$.
(For the isotropic 3D or 2D materials considered here, $\boldsymbol{\sigma}$
reduces to the scalar $\sigma = \Tr\boldsymbol{\sigma} / d$, for which
$\vec{v}\otimes\vec{v}$ above can be replaced by $v^2/d$.)
In turn, the momentum relaxation rate for each electronic state is evaluated using Fermi's rule,
\begin{align}
(\tau^p_{\vec{k}n})^{-1} =
\frac{2\pi}{\hbar} \int\sub{BZ} \frac{\Omega \,\D\vec{k}'}{(2\pi)^d} &\sum_{n'\alpha\pm}
	\delta(\varepsilon_{\vec{k}'n'} - \varepsilon_{\vec{k}n} \mp \hbar\omega_{\vec{k}'-\vec{k},\alpha})
\nonumber \\
\times
	&\left( n_{\vec{k}'-\vec{k},\alpha} + \frac{1}{2} \mp \left(\frac{1}{2} - f_{\vec{k}'n'}\right)\right)
	\left| g^{\vec{k}'-\vec{k},\alpha}_{\vec{k}'n',\vec{k}n} \right|^2 
\nonumber \\
\times
	&\left(1 - \frac{\vec{v}_{\vec{k}n}\cdot\vec{v}_{\vec{k}'n'}}{|\vec{v}_{\vec{k}n}| |\vec{v}_{\vec{k}'n'}|}\right),
\label{eqn:tauInv_ePhP}
\end{align}
where $\omega_{\vec{q}\alpha}$ and $n_{\vec{q}\alpha}$ are energies
and Bose occupation factors of phonons with wave-vector $\vec{q}$
($=\vec{k}'-\vec{k}$ above by momentum conservation) and polarization index
$\alpha$, $\Omega$ is the unit cell volume (or area when $d=2$),
$g^{\vec{k}'-\vec{k},\alpha}_{\vec{k}'n',\vec{k}n}$ are the electron-phonon
matrix elements and the sum over $\pm$ accounts for phonon absorption and emission.
The final factor accounts for the scattering angle in the momentum relaxation rate.
We also report the average momentum relaxation time,
\begin{equation}
\tau_{{\mathrm{D}}0} =\frac{\displaystyle
\int\sub{BZ}\frac{g\sub{s}\,\D\vec{k}}{(2\pi)^d} \sum_n 
	\frac{\partial f_{\vec{k}n}}{\partial\varepsilon_{\vec{k}n}} |\vec{v}_{\vec{k}n}|^2 \tau_{\vec{k}n}^p
 }{\displaystyle
\int\sub{BZ}\frac{g\sub{s}\,\D\vec{k}}{(2\pi)^d} \sum_n 
	\frac{\partial f_{\vec{k}n}}{\partial\varepsilon_{\vec{k}n}} |\vec{v}_{\vec{k}n}|^2
},
\label{eqn:tauD0}
\end{equation}
where the weight factors reflect the relative contributions
of various electronic states to the conductivity.

\subsection*{Optical response}

The optical response can be expressed in terms of the AC conductivity
\begin{equation}
\boldsymbol{\sigma}(\omega) = \frac{\boldsymbol{\sigma}_0\tau_{{\mathrm{D}}0}^{-1}}{\tau_{\mathrm{D}}^{-1}(\omega) + \iu\omega} + \boldsymbol{\sigma}_{\mathrm{d}}(\omega),
\label{eqn:sigmaAC}
\end{equation}
where the first term captures the Drude response including 
the effect of phonon-assisted intraband transitions,
while the second term captures the effect of direct optical transitions.
We evaluate the second term directly using Fermi's golden rule
for the real part (imaginary part of corresponding $\epsilon(\omega)$),\cite{DirectTransitions}
\begin{multline}
\Re\boldsymbol{\sigma}_{\mathrm{d}}(\omega)
= \frac{\pi e^2}{\epsilon_0\omega} \int\sub{BZ} \frac{g\sub{s}\,\D\vec{k}}{(2\pi)^d} \sum_{n'n}
	(f_{\vec{k}n} - f_{\vec{k}n'})
\\\times
	\delta(\varepsilon_{\vec{k}n'} - \varepsilon_{\vec{k}n} - \hbar\omega)
	\left(\vec{v}^{\vec{k}\ast}_{n'n} \otimes \vec{v}^{\vec{k}}_{n'n}\right),
\label{eqn:sigmaDirect}
\end{multline}
where $\vec{v}^{\vec{k}}_{n'n}$ are the matrix-elements of the velocity operator.
We then evaluate the imaginary part from it using the Kramers-Kronig relation.
Above the energy-conserving $\delta$-function is broadened to a Lorentzian
due to carrier linewidths from electron-electron and electron-phonon scattering,
which we also calculate using the same first-principles framework.\cite{PhononAssisted}

In the first term above, we capture the intraband response 
including phonon-assisted transitions by evaluating the
frequency-dependent momentum relaxation rate from
the Eliashberg spectral function,\cite{AllenEliashberg}
generalized here to finite temperature as
\begin{equation}
\tau_{\mathrm{D}}^{-1}(\omega) = \frac{2\pi}{\hbar g(\varepsilon_{\mathrm{F}}) b_T(\hbar\omega)}
	\sum_\alpha\int\sub{BZ} \frac{\D\vec{q}}{(2\pi)^d}
		G^p_{\vec{q}\alpha} b_T(\hbar\omega-\hbar\omega_{\vec{q}\alpha}).\nonumber
\end{equation}
Above $g(\varepsilon_{\mathrm{F}})$ is the density of electronic states at the Fermi level, and
we define $b_T(\varepsilon) \equiv \varepsilon/(1-\mathrm{e}^{-\varepsilon/k_{\mathrm{B}}T})$ and the dimensionless
\begin{multline}
G^p_{\vec{q}\alpha} \equiv \sum_{nn'}\int\sub{BZ}\frac{g\sub{s}\Omega\,\D\vec{k}}{(2\pi)^d}
	\left| g^{\vec{q}\alpha}_{(\vec{k}+\vec{q})n',\vec{k}n} \right|^2
	\left(1 - \frac{\vec{v}_{\vec{k}n}\cdot\vec{v}_{(\vec{k}+\vec{q})n'}}{|\vec{v}_{\vec{k}n}| |\vec{v}_{(\vec{k}+\vec{q})n'}|}\right)
\\\times
	\delta(\varepsilon_{\vec{k}n}-\varepsilon_{\mathrm{F}})
	\delta(\varepsilon_{(\vec{k}+\vec{q})n'}-\varepsilon_{\mathrm{F}}),
\end{multline}
which represents the total coupling of each phonon mode to electronic states near the Fermi level.
($G^p_{\vec{q}\alpha}$ is the weight of a phonon mode in the `transport
Eliashberg spectral function', which accounts for momentum scattering angle
compared to the conventional Eliashberg spectral function.\cite{AllenEliashberg})
Finally, the numerator in the first term of Eq.~\eqref{eqn:sigmaAC}
is effectively the Fermi-surface-integrated square velocity,
\begin{equation}
\frac{\boldsymbol{\sigma}_0}{\tau_{\mathrm{D}0}} = \int\sub{BZ}\frac{e^2g\sub{s}\,\D\vec{k}}{(2\pi)^d} \sum_n
	\delta(\varepsilon_{\vec{k}n} - \varepsilon_{\mathrm{F}}) (\vec{v}_{\vec{k}n}\otimes\vec{v}_{\vec{k}n}),
\end{equation}
essentially the full-bands generalization of the term $g(\varepsilon_{\mathrm{F}})v_{\mathrm{F}}^2/d$
that appears in the Drude theory of $d$-dimensional metals.

\subsection*{Plasmonic properties}

\paragraph*{2D layers}
The optical response of a 2D layer is dictated by the frequency-dependent 2D conductivity $\sigma\sub{2D}(\omega)$: 
it links the induced surface current $\mathbf{K}$ linearly to the (total) in-plane electric field $\mathbf{E}_{\parallel}$.
Paired with Maxwell's equations, this constitutive relation is sufficient to analyze the properties of any 2D polaritons, 2D plasmons included.
If a plasmon exists, it manifests as a pole in the monolayer's TM reflection coefficient~\cite{Koppens:2011, Bludov:2013, ChristensenThesis:2017}
\begin{equation}
r\sub{tm}(q,\omega) = \frac{q_\perp\sigma\sub{2D}(\omega)}{2\varepsilon_0\omega + q_\perp\sigma\sub{2D}(\omega)},
\end{equation}
with in-plane, out-of-plane, and free-space wave vectors $q$, $q_\perp^2\equiv k_0^2-q^2$,
and $k_0\equiv \omega/c$, respectively.
The poles dictate the plasmon dispersion equation, $q = [(2\iu\varepsilon_0\omega/\sigma\sub{2D})^2 + k_0^2]^{1/2}$.

\vskip 1ex
\paragraph*{Finite slabs}
Our considerations of finite Ag slabs (defined as slab-like for thicknesses $t>t\sub{2D}$), employ the bulk dielectric function $\epsilon(\omega)$ of Ag. 
The TM reflection coefficient of the vacuum-clad slab is computed from standard formula, see e.g. Ref.~\citenum{novotnyhecht:2012}.
For a metallic slab, the associated TM reflection coefficient exhibits two distinct pole species, reflecting the existence of two plasmonic branches: 
one low-energy branch (associated with a charge-even mode, or, equivalently an odd $\mathbf{H}$-field) and a high-energy branch (associated with a charge-odd mode, equivalently an even $\mathbf{H}$-field). In the limit of vanishing thickness, $t\rightarrow 0$, the charge-even mode asymptotically approaches the 2D layer's dispersion: thus, this is the mode of interest for comparisons with 2D plasmonics (consequently,  the charge-odd mode is omitted here).
Its dispersion equation is~\cite{Economou:1969}:
\begin{equation}\label{eq:slabdisp_evenmode}
\coth\left(\frac{-\iu q_\perp^\prime t}{2} \right) = -\frac{\epsilon(\omega)q_\perp}{q_\perp^\prime},
\end{equation}
with out-of-plane wave vectors $q_\perp^2\equiv k_0^2-q^2$ and $(q_\perp^\prime)^2\equiv \varepsilon(\omega) k_0^2-q^2 $ associated with the vacuum-cladding and slab-regions, respectively. Equation~\eqref{eq:slabdisp_evenmode} is a transcendental equation; in practice, we solve it by numerical minimization.

\bibliographystyle{apsrev4-1}
\bibliography{references}

\end{document}